# A VLBI Resolution of the Pleiades Distance Controversy

Carl Melis[1*], Mark J. Reid[2], Amy J. Mioduszewski[3], John R. Stauffer[4], Geoffrey C. Bower[5]

[1]University of California, San Diego, Center for Astrophysics and Space Sciences, 9500 Gilman Dr., La Jolla, CA 92093-0424, USA.

[2]Harvard-Smithsonian Center for Astrophysics, 60 Garden St, Cambridge, MA 02138, USA.

[3]National Radio Astronomy Observatory, Array Operations Center, 1003 Lopezville Road, Socorro, NM 87801, USA.

[4]Spitzer Science Center (SSC), 1200 E. California Blvd., California Institute of Technology, Pasadena, CA 91125, USA.

[5]Academia Sinica Institute of Astronomy and Astrophysics (ASIAA), 645 N. A'ohoku Pl., Hilo, HI 96720, USA.

*To whom correspondence should be addressed; E-mail: cmelis@ucsd.edu.

**Because of its proximity and its youth, the Pleiades open cluster of stars has been extensively studied and serves as a cornerstone for our understanding of the physical properties of young stars. This role is called into question by the "Pleiades distance controversy" wherein the cluster distance of 120.2±1.5 pc as measured by the optical space astrometry mission *Hipparcos* is significantly different from the distance of 133.5±1.2 pc derived with other techniques. We present an absolute trigonometric parallax distance measurement to the Pleiades cluster that uses very long baseline radio interferometry. This distance of 136.2±1.2 pc is the most accurate and precise yet presented for the cluster and is incompatible with the *Hipparcos* distance determination. Our results cement existing astrophysical models for Pleiades-age stars.**

Robust physical parameters for stars can only be obtained when an estimate of the distance to the object of interest exists. Trigonometric parallax – which uses the orbit of the Earth around the Sun to inform the principle of triangulation – provides the most fundamental distance measurement outside of our Solar system. High-precision tests of stellar physical models thus rely heavily on collections of parallax determinations. With reasonable physical models for nearby stars – and some mild assumptions about the homogeneity of classes of astrophysical objects throughout the Universe (the Vogt-Russell theorem; see e.g., refs. *1,2*) – distance estimates for sources that lie beyond the current limit of trigonometric parallax can be systematically compiled. Such a methodology forms the basis of the cosmic distance ladder that elucidates the structure and evolution of the Universe (*3*).

Clusters of coeval stars yield a solid foundation for tests of stellar physical models. Young open clusters are especially important because their stellar constituents define the "zero-age main sequence" – the curve along which stable, core-hydrogen burning stars reside in a color-magnitude diagram. Empirical isochrones developed from these young open clusters can be applied to other vastly more distant groups of stars (when brightness measurements of individual stars in the group can be made) to estimate their distance, thus providing structural information for the galaxies that contain them (*4,5*). The Pleiades open cluster of stars is critical for such studies because its relatively young age places many of its stars on the zero-age main sequence. It is the closest cluster to Earth of its age and richness of stars and thus lends itself to highly detailed investigations. One would expect that all astrophysical parameters for such an important sample of stars would be well characterized. However, there still rages an open debate regarding the distance to the Pleiades.

Figure 1 summarizes distances obtained for the Pleiades cluster to date, including the new measurement described here. As can be seen, most measurements are in rough agreement with that produced in this work, with the stark exception of the *Hipparcos* astrometric satellite distances. For a single object near the distance of the Pleiades, *Hipparcos* was not capable of producing a distance measurement with accuracy better than 10%. However, by taking the aggregate of many cluster members, *Hipparcos* was able to achieve a Pleiades parallax with roughly 1% precision (*6,7*). In almost any other case, one would simply discard the disagreeable *Hipparcos* cluster distances as bad measurements, but the *Hipparcos* mission represents the most complete astrometric survey of the sky and of the Pleiades cluster to date. It provides a path that is free of stellar physical models to obtaining the cluster distance and combines more than 50 cluster member distance measurements. Other methods either include at most several cluster members in their distance determination, rely heavily on physical models to obtain a cluster distance (whereas it should be the distance measurement that informs the development of physical models), or result in large uncertainties in the cluster distance.

Although the discrepancy between *Hipparcos* and the average non-*Hipparcos* distance (Fig. 1) amounts to a 10% difference, the resultant changes to physical models needed to obtain agreement with the *Hipparcos* value are quite significant. One such change requires a 20 to 40% increase in the amount of helium (He) that Pleiades stars are composed of (*5*), a change that throws into question any attempt to systematically apply model isochrones to groups of stars that have not been characterized in great detail because one typically only has brightness measurements at a few wavelengths (Making compositional measurements is extremely resource expensive and He measurements in particular are difficult. He measurements to date suggest that stars formed in the recent Galactic history have similar He abundances) A more disconcerting explanation invokes different, unknown physics for young stars of roughly Pleiades-age (*6*), thus challenging our general understanding of the star formation and evolution process. As a result, the controversy surrounding the distance to the Pleiades has not subsided. On the contrary, it has grown as each side of the debate has exchanged their own views, and neither side has backed down (*7,8*).

Given the disagreement between parallax measurements using a similar methodology (relative astrometry in the optical wavelengths), we pursued a new approach that could provide an independent view on Pleiades cluster distance measurements made to date. Our approach uses radio astrometry (*9*), a technique that provides an absolute distance measurement via referencing to an essentially stationary (to within our measurement capabilities) quasi-stellar object (an actively accreting supermassive blackhole in the distant Universe). To achieve sufficient precision (better than 0.0001 seconds of arc) in stellar position measurements, we made observations using an array of widely-separated radio antennas that when acting in concert give the resolution of a telescope the size of Earth. The very long baseline interferometry (VLBI) array employed by our study uses the Very Long Baseline Array (VLBA) as its core and additionally incorporates the Robert C. Byrd Green Bank Telescope, the Effelsberg Radio Telescope, and the William E. Gordon Telescope at Arecibo Observatory for enhanced resolution and sensitivity. Four Pleiades star systems were observed with this array over a period of ~1.5 years to completely map their parallax motion (Supplementary Materials, Tables S1-S5, and Fig. 2). Model fits to the motion of each star on the plane of the sky produce the desired parallax measurement (Table 1). The measured distances and ±1 SD errors for the four systems are 134.8±0.5 pc (HII 174), 138.4±1.1 pc (HII 625), 135.5±0.6 pc (HII 1136), and 136.6±0.6 pc (HII 2147 system). Of note is the <1% accuracy for the individual object VLBI distance measurements.

Already evident in each individual stellar distance measurement for our sample is gross disagreement with the *Hipparcos* cluster distance. To derive the cluster absolute parallax, however, one must include with the measurements of the individual stars the additional uncertainty of each star's position with respect to the center of the cluster. We adopt the approach of Soderblom *et al*. (*8*) of using the 1σ angular dispersion of the cluster as the systematic cluster depth uncertainty. For an assumed Pleiades distance of 130 pc and cluster dispersion of 1°, we estimate the cluster depth uncertainty to be 2.3 pc and add this value in quadrature to each object's formal distance uncertainty. This additional error component dominates the final cluster distance uncertainty. When averaging individual VLBI measured distances to arrive at the final cluster distance, we treat HII 2147 as a single system and use the average of the distance measurements for HII 2147 NE and SW as given above. In this way, we calculate the VLBI-measured Pleiades cluster distance to be 136.2±1.2 pc (±1 SD).

An important aspect of this independent VLBI distance measurement is that it validates previous non-*Hipparcos* parallax and binary orbit distance measurements. As such, we can combine all parallax (including VLBI) and binary orbit distances into a single non-*Hipparcos* cluster distance; this sample includes 17 individual Pleiades star systems. Due to their fitting techniques which result in coupled individual Pleiades member parallaxes, we treat each of the distance measurements of Soderblom *et al.* (*8*) and Gatewood *et al.* (*10*) as a single system measurement similar to the case of HII 2147 above. Each of the VLBI individual parallaxes, the two binary orbit distances (*11,12*), and the distances of Soderblom *et al.* (*8*) and Gatewood *et al.* (*10*) are combined with a weighted mean. In deriving the combined cluster distance and associated uncertainty, cluster depth uncertainty is added in quadrature to the uncertainty of each system distance measurement. From this, we obtain a non-*Hipparcos* Pleiades cluster distance of 136.1±1.0 pc (the vertical grey band in Fig. 1; this value is nearly identical to the VLBI-measured cluster distance because the VLBI parallaxes have the smallest uncertainty and hence carry the most weight).

Our results conclusively show that the *Hipparcos*-measured distance to the Pleiades cluster is in error. The general agreement of our distance measurement with those distances obtained by isochrone fitting in Fig. 1 suggest that physical models provide an accurate representation of the properties of Pleiades-age stars and that no unusual compositions or unknown physics lurk within this canonical cluster. Although this is likely a great relief for modelers of stars, it raises further questions into what happened with *Hipparcos*. Whatever error that manifested itself as a significantly skewed distance to the Pleiades cluster remains at large (some have suggested possible explanations, see e.g., refs *13,14*). The unrecognized nature of such an error is especially dangerous when one considers that *Gaia* (*15*) – the successor to *Hipparcos* and very similar in design – is just now starting its Galaxy-mapping mission. If the unrecognized *Hipparcos* error has crept into the *Gaia* pipeline, how would it manifest itself (if it does)? VLBI distance measurements like those presented here will serve as an important cross-check of the *Gaia* output near its predicted precision limits.

We thank the National Radio Astronomy Observatory, Green Bank Telescope, Arecibo Observatory, and Effelsberg Telescope staff who coordinated, conducted, and correlated observations for this project. All data presented in this paper are maintained in the National Radio Astronomy Observatory archive. The National Radio Astronomy Observatory is a facility of the National Science Foundation operated under cooperative agreement by Associated Universities, Inc. This work made use of the Swinburne University of Technology software correlator, developed as part of the Australian Major National Research Facilities Programme and operated under license. C. M. acknowledges financial support from the US National Science Foundation through awards AST-1313428 and AST-1003318, from a Lawrence Livermore National Laboratory minigrant to UCLA, and from the Spitzer Science Center Visiting Graduate Student Program. G. C. B. acknowledges support from the Academica Sinica Institute for Astronomy and Astrophysics.

**Fig. 1. Pleiades cluster distances.** Summary of Pleiades distances obtained through various methods. The red asterisk with a distance of 136.2±1.2 pc is the new VLBI determination. The blue triangles near 120 pc are from two reductions of the *Hipparcos* data (*6,7*). The vertical dashed line with uncertainty range shown by dotted lines and filled in with gray is the cluster distance derived from non-*Hipparcos* trigonometric parallaxes and binary orbits as described in the text. All plotted errors are ±1 SD. References for the distances shown, from top to bottom for each category, are as follows: Isochrone Fitting – An *et al.* (*4*), Percival *et al.* (*19*), Stello & Nissen (*20*), Pinsonneault *et al.* (*5*), Giannuzzi (*21*), van Leeuwen (*22*), Nicolet (*23*); Trigonometric Parallax (excluding *Hipparcos* and VLBI) – Soderblom *et al.* (*8*), Gatewood *et al.* (*10*); Orbital Modeling – Groenewegen *et al.* (*11*), Southworth *et al.* (*24*), Zwahlen *et al.* (*12*), Munari *et al.* (*25*), Pan *et al.* (*26*); Moving Cluster – Röser & Schilbach (*27*), Narayanan & Gould (*13*).

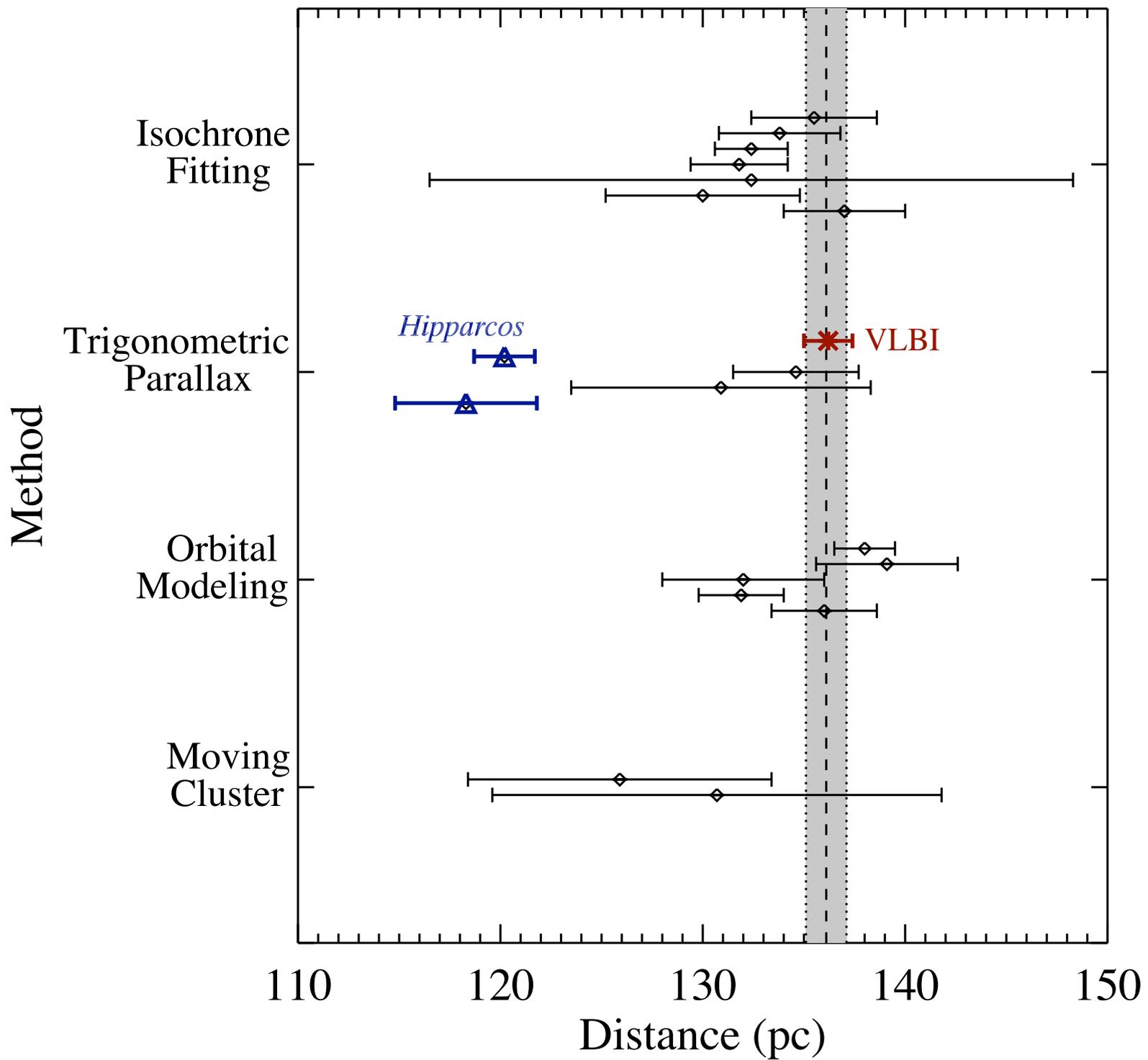

**Fig. 2. VLBI Pleiad parallaxes.** Parallax fits to VLBI position measurements and associated random errors (±1 SD) for five Pleiades stars, including both components of the HII 2147 binary system. For each object the solid line is the best-fitting astrometric model that includes proper motion and parallax; the proper motion has been removed in the data points to accentuate the parallax motion. For each component of the HII 2147 binary system and for HII 1136 we additionally include acceleration terms in our fit to model short segments of a binary orbit (the average angular separation between the two stars of the HII 2147 system over the monitoring period reported in Tables S4 and S5 is ≈60 milliarcseconds or ≈8.2 AU in projection). The left-hand panel curves and data points show East (right ascension times cos(declination)) angular offsets on the sky of the source position relative to a distant quasar. The right-hand panel curves and data points show North (declination) offsets. Each source is color-coded and labeled in the declination panels.

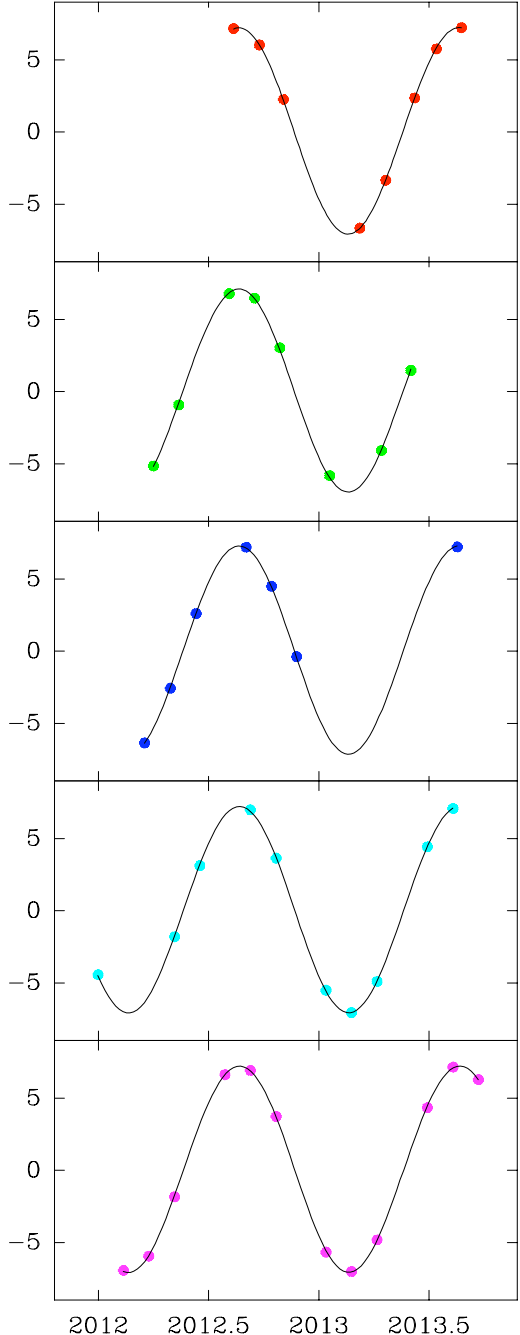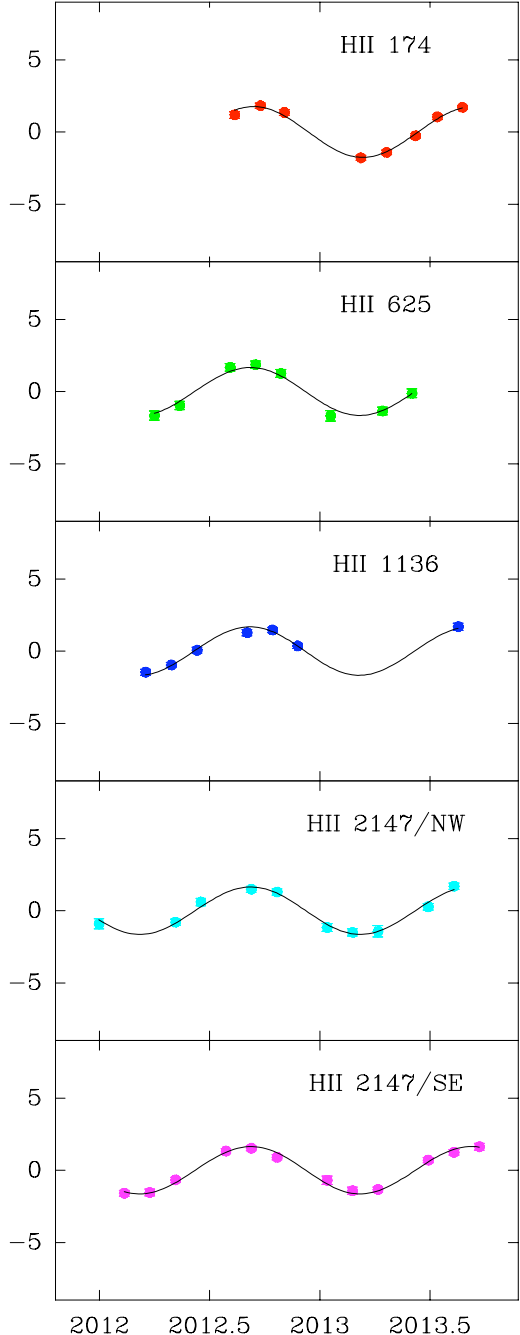

| Source Name | Optical Proper Motions (17,18) | | Fitted Parameters | | | | | |
|---|---|---|---|---|---|---|---|---|
| | $\mu_\alpha\cos\delta$ | $\mu_\delta$ | $\pi$ | $\mu_\alpha\cos\delta$ | $\mu_\delta$ | $a_\alpha\cos\delta$ | $a_\delta$ | $\chi^2$ |
| | (mas yr$^{-1}$) | (mas yr$^{-1}$) | (mas) | (mas yr$^{-1}$) | (mas yr$^{-1}$) | (mas yr$^{-2}$) | (mas yr$^{-2}$) | |
| HII 174 | 22.0±2.0 | -45.7±2.1 | 7.418±0.025 | 19.86±0.05 | -45.41±0.16 | – | – | 1.018 |
| HII 625 | 20.0±2.0 | -47.9±6.9 | 7.223±0.057 | 19.47±0.11 | -44.39±0.27 | – | – | 1.002 |
| HII 1136 | 17.3±0.7 | -44.8±1.8 | 7.382±0.031 | 17.18±0.05 | -47.39±0.24 | -0.43±0.16 | 0.6±0.8 | 0.941 |
| HII 2147 NW | 17.1±1.0 | -45.4±0.7 | 7.328±0.035 | 23.22±0.05 | -46.76±0.16 | 1.73±0.19 | -3.9±0.7 | 1.008 |
| HII 2147 SE | | | 7.319±0.027 | 14.05±0.04 | -42.24±0.11 | -1.05±0.18 | 2.0±0.5 | 0.982 |

**Table 1. Fitted astrometric parameters.** For each object in our sample, we conducted astrometric fits to the measured positions to extract stellar parallaxes. Only data taken in 2012-2013 were used for HII 1136 to ensure consistent and readily comparable results. Measured positions are modeled with the sum of a parallax sinusoid (determined by the parallax magnitude – $\pi$ – and the purely geometrical motion for a given part of the sky induced by Earth's orbit), a reference position at an arbitrarily chosen fixed epoch, and a linear or accelerated proper motion ($\mu_\alpha\cos\delta$, $\mu_\delta$, $a_\alpha\cos\delta$, and $a_\delta$; acceleration terms account for binary motion when the orbital period is much larger than the time frame over which the system was monitored and have been successfully used in past attempts to measure system parallaxes, see ref. *16*). This results in five or seven fitted model parameters. During the fitting process the data are weighted using the quadrature sum of the formal measured fit uncertainties and an additional component that represents systematic uncertainties. A least squares fitting routine determines the parameters that minimize the sum of the squares of the residuals. This process allows the systematic error component to be adjusted as necessary to obtain a $\chi^2$ equal to 1 for each of the R.A. and Decl. data. The fitted proper motions can be compared with the values shown in second and third columns that were previously determined from optical measurements. All uncertainties are ±1 SD.

**Supplementary Text**

In this section some comments on the target sample are given and the observations are discussed in detail.

Target Sample

The target systems were selected from lists of well-characterized Pleiades stars and thus there is little doubt that they are members of the cluster. Indeed, the Pleiades proper motion signature is robust and the previously measured optical proper motion values for each system by themselves identify all systems as high probability cluster members (*28*). Inclusion in the VLBI program was determined from an initial radio-imaging survey by our team that targeted rapidly rotating and X-ray luminous Pleiades members (*29* – several other Pleiades members identified by ref. *29* to be radio-loud are currently under investigation with VLBI but do not yet have final results). It is worth noting that none of our target systems are included in the *Hipparcos* Pleiades sample as they are too faint in the optical (visual magnitudes and colors are given for each system in Tables S1-S5).

Summary of the Observations and Datasets

Observations of the four Pleiades star systems were conducted with a very long baseline radio interferometer consisting of the Very Long Baseline Array (VLBA, which consists of 10 identical 25-m antennas in Mauna Kea HI, Brewster WA, Owens Valley CA, Kitt Peak AZ, Pie Town NM, Los Alamos NM, Fort Davis TX, North Liberty IA, Hancock MA, and Saint Croix Virgin Islands), the 100-m Robert C. Byrd Green Bank Telescope in West Virginia, the Effelsberg 100-m Radio Telescope in Bad Münstereifel Germany, and the 300-m William E. Gordon Telescope at Arecibo Observatory in Puerto Rico. Baseline lengths for these antennas range from a minimum of 236 km to a maximum of 10,328 km. It is worth noting that not all 13 stations were available for every epoch and that not every station produced useful data in each epoch. Preliminary observations of one system, HII 1136, began in 2004 and continued through 2010. The main program observations for the sources in Tables S1-S5 began in late 2011 and ended in 2013. Tables S1-S5 list specific observation dates for each system.

Each system was observed in continuum light centered at a frequency of 8.4 GHz (roughly 3.6 cm). During this project the VLBA was undergoing upgrades that enabled wider bandwidth observations, and thus the precise average continuum frequency changed with the bandwidth used. Tables S1-S5 list instrumental setups for each observing epoch. The background quasar J0347+2339 (which has a measured R.A. of 03h47m57.11171s ± 1.3 mas and Decl. of +23°39'55.3248" ± 2.2 mas) served as the main phase-reference source as its typical separation from our Pleiades targets was less than 1° (separations of each target from this reference source are given in Tables S1-S5). Because of the fortuitous placement of J0347+2339 with respect to our target stars, and the intrinsic faintness for most targets, we did not pursue geodetic observations during tracks – geodetic observations often improve astrometric accuracy when using wider separation reference sources (*30*). Observation tracks consisted of scans on bright background quasars that are used to set the instrumental delays and a series of cycles where roughly 1.5 minutes were spent on the target star and roughly 1 minute was spent on the phase-reference source. Although each track was of 10 hours in duration, typically only 4-6 hours of that time were spent on the target source.

Data were recorded on hard disks at each station, then mailed to the Pete V. Domenici Science Operations Center (SOC) in Socorro, NM for correlation. For epochs before 2010.6, a hardware correlator was employed. After that time, correlation was done with the software correlator developed by Deller *et al.* (*31*). Correlated data sets were retrieved from the NRAO data archive service through the world wide web.

Data reduction follows standard phase-referenced radio interferometry practices for very long baseline astrometry datasets. Editing and calibration of each dataset occurrs within the Astronomical Image Processing System (AIPS; *32*) software suite. Earth-orientation parameters for each observation date are obtained from the US Naval Observatory database and applied to the data to correct estimated values utilized by the VLBA correlator. Dispersive delays to incoming radio light caused by free electrons in the Earth's atmosphere are accounted for through the use of an estimate of the electron content of the ionosphere derived from Global Positioning System (GPS) measurements. Amplitude calibration for each antenna is then obtained from measured system temperatures and standard gain curves. Phase corrections due to antenna parallactic angle effects are then applied. Instrumental delays are removed for each antenna and spectral sub-band by fringe fitting a single strong calibration source (typically the quasar J0403+2600). A final, global fringe-fitting pass is made for the main phase-reference source under the assumption that it is point like for all antenna pairs. This final calibration is then applied to the target star scans.

Once calibrated, the target source visibility data are imaged on a spatial grid with pixel size of 110 µas. The map rms noise level obtained is a strong function of which antennas produced useable data in each epoch. These noise levels are listed for each epoch in Tables S1-S5. The variable nature of the emission from the target sources led to numerous epochs with no detections. When detected, the absolute position of the targets is obtained from a two-dimensional Gaussian fitting procedure. Errors associated with these fits are also obtained based on the expected theoretical astrometric performance of an interferometer (*33*). However, systematic errors from uncompensated tropospheric and ionospheric delays also contribute to the uncertainty in source position and their contribution is quantified during the parallax fitting process. It is worth noting that the uncertainties for the absolute positions given in Tables S1-S5 do not include the additional error on the absolute position of the primary phase reference source given previously. However, when performing astrometric fits (to extract parallax and proper motion) it is only the motion of each target relative to the stationary background quasar that needs to be considered.

**Table S1.**
Observations Summary for HII 174

| Separation from J0347+2339 = 1.638° | | | | | $V_{mag}$= 11.6, $B_{mag}$–$V_{mag}$ color = 0.85 | | | |
|---|---|---|---|---|---|---|---|---|
| Observation Date (mid-track) | Band width | Central Frequency | Measured R. A. | R. A. error (one S.D.) | Measured Decl. | Decl. error (one S.D.) | Flux | Map rms noise |
| Julian Date Years | MHz | GHz | h m s | s | ° ′ ″ | ″ | μJy | μJy |
| 2455940.57 2012.0357 | 128 | 8432.9 | — | — | — | — | — | 19 |
| 2455982.46 2012.1502 | 128 | 8432.9 | — | — | — | — | — | 19 |
| 2456025.34 2012.2673 | 128 | 8432.9 | — | — | — | — | — | 19 |
| 2456067.23 2012.3818 | 128 | 8432.9 | — | — | — | — | — | 101 |
| 2456109.11 2012.4962 | 128 | 8432.9 | — | — | — | — | — | 33 |
| 2456150.99 2012.6106 | 128 | 8432.9 | 03 43 48.3513620 | 0.0000060 | +25 00 15.221939 | 0.000192 | 173 | 24 |
| 2456193.91 2012.7279 | 128 | 8432.9 | 03 43 48.3514447 | 0.0000022 | +25 00 15.217191 | 0.000083 | 66 | 16 |
| 2456233.77 2012.8368 | 128 | 8432.9 | 03 43 48.3513229 | 0.0000051 | +25 00 15.211728 | 0.000128 | 108 | 19 |
| 2456275.66 2012.9513 | 512 | 8415.9 | — | — | — | — | — | 10 |
| 2456318.54 2013.0686 | 512 | 8415.9 | — | — | — | — | — | 9 |
| 2456360.43 2013.1834 | 512 | 8415.9 | 03 43 48.3511691 | 0.0000027 | +25 00 15.192795 | 0.000073 | 72 | 9 |
| 2456403.31 2013.3009 | 512 | 8415.9 | 03 43 48.3515858 | 0.0000018 | +25 00 15.187847 | 0.000054 | 141 | 12 |
| 2456451.18 2013.4320 | 512 | 8415.9 | 03 43 48.3521994 | 0.0000015 | +25 00 15.183080 | 0.000045 | 190 | 10 |

| 2456487.08 2013.5304 | 512 | 8416.2 | 03 43 48.3525964 | 0.0000025 | +25 00 15.179952 | 0.000104 | 90 | 15 |
| 2456528.97 2013.6451 | 512 | 8415.9 | 03 43 48.3528768 | 0.0000015 | +25 00 15.175447 | 0.000046 | 96 | 9 |

**Table S2.**

Observations Summary for HII 625

| Separation from J0347+2339 = 0.598° | | | | $V_{mag}$= 12.7, $B_{mag}$–$V_{mag}$ color = 1.2 | | | | |
|---|---|---|---|---|---|---|---|---|
| Observation Date (mid-track) | Band width | Central Frequency | Measured R. A. | R. A. error (one S.D.) | Measured Decl. | Decl. error (one S.D.) | Flux | Map rms noise |
| Julian Date Years | MHz | GHz | h m s | s | ° ' " | " | µJy | µJy |
| 2455933.59 2012.0167 | 128 | 8432.9 | — | — | — | — | — | 19 |
| 2455975.48 2012.1311 | 128 | 8432.9 | — | — | — | — | — | 17 |
| 2456018.36 2012.2483 | 128 | 8432.9 | 03 45 21.2035163 | 0.0000042 | +23 43 38.340643 | 0.000179 | 111 | 14 |
| 2456060.25 2012.3627 | 128 | 8432.9 | 03 45 21.2039877 | 0.0000072 | +23 43 38.336247 | 0.000143 | 346 | 41 |
| 2456102.14 2012.4772 | 128 | 8432.9 | — | — | — | — | — | 18 |
| 2456144.01 2012.5916 | 128 | 8432.9 | 03 45 21.2048741 | 0.0000025 | +23 43 38.328707 | 0.000068 | 112 | 14 |
| 2456185.93 2012.7061 | 128 | 8432.9 | 03 45 21.2050142 | 0.0000009 | +23 43 38.323798 | 0.000027 | 439 | 16 |
| 2456227.79 2012.8205 | 128 | 8432.9 | 03 45 21.2049262 | 0.0000040 | +23 43 38.318108 | 0.000077 | 70 | 14 |
| 2456269.68 2012.9349 | 128 | 8432.9 | — | — | — | — | — | 18 |
| 2456310.56 2013.0468 | 128 | 8432.9 | 03 45 21.2046021 | 0.0000093 | +23 43 38.305108 | 0.000249 | 98 | 23 |
| 2456353.45 2013.1642 | 128 | 8432.9 | — | — | — | — | — | 16 |
| 2456396.32 2013.2817 | 128 | 8432.9 | 03 45 21.2050618 | 0.0000059 | +23 43 38.295017 | 0.000099 | 110 | 21 |
| 2456445.27 2013.4158 | 512 | 8415.9 | 03 45 21.2056562 | 0.0000081 | +23 43 38.290298 | 0.000182 | 94 | 22 |

**Table S3.**

Observations Summary for HII 1136

| Separation from J0347+2339 = 0.338° | | | | | $V_{mag}$= 12.2, $B_{mag}$–$V_{mag}$ color = 1.0 | | | |
|---|---|---|---|---|---|---|---|---|
| Observation Date (mid-track) | Band width | Central Frequency | Measured R. A. | R. A. error (one S.D.) | Measured Decl. | Decl. error (one S.D.) | Flux | Map rms noise |
| Julian Date Years | MHz | GHz | h m s | s | ° ‘ “ | “ | μJy | μJy |
| 2453085.35 2004.2182 | 64 | 8421.5 | 03 46 40.2509284 | 0.0000048 | +23 29 51.676625 | 0.000138 | 284 | 30 |
| 2453086.35 2004.2209 | 64 | 8421.5 | 03 46 40.2509467 | 0.0000056 | +23 29 51.676556 | 0.000203 | 165 | 29 |
| 2453408.55 2005.1015 | 64 | 8421.5 | — | — | — | — | — | 44 |
| 2453456.42 2005.2327 | 64 | 8421.5 | — | — | — | — | — | 41 |
| 2453526.21 2005.4239 | 64 | 8421.5 | 03 46 40.2530800 | 0.0000080 | +23 29 51.622538 | 0.000331 | 363 | 34 |
| 2453588.05 2005.5933 | 64 | 8421.5 | — | — | — | — | — | 33 |
| 2455268.46 2010.1944 | 128 | 8432.9 | 03 46 40.2587312 | 0.0000026 | +23 29 51.398328 | 0.000094 | 108 | 23 |
| 2455423.01 2010.6178 | 128 | 8432.9 | — | — | — | — | — | 23 |
| 2455429.99 2010.6370 | 128 | 8432.9 | 03 46 40.2602852 | 0.0000057 | +23 29 51.380726 | 0.000155 | 93 | 26 |
| 2455606.50 2011.1205 | 128 | 8432.9 | — | — | — | — | — | 20 |
| 2455919.64 2011.9785 | 128 | 8432.9 | — | — | — | — | — | 23 |
| 2455961.52 2012.0930 | 128 | 8432.9 | — | — | — | — | — | 15 |
| 2456003.41 2012.2074 | 128 | 8432.9 | 03 46 40.26131586 | 0.0000037 | +23 29 51.3039153 | 0.000115 | 196 | 15 |
| 2456046.29 2012.3246 | 128 | 8432.9 | 03 46 40.26173818 | 0.0000013 | +23 29 51.2988673 | 0.000050 | 273 | 15 |
| 2456089.18 2012.4417 | 128 | 8432.9 | 03 46 40.26226048 | 0.0000041 | +23 29 51.2943494 | 0.000085 | 259 | 15 |
| 2456130.05 2012.5534 | 128 | 8432.9 | — | — | — | — | — | 15 |
| 2456171.95 2012.6679 | 128 | 8432.9 | 03 46 40.2628778 | 0.0000034 | +23 29 51.284851 | 0.000124 | 87 | 15 |
| 2456213.82 2012.7823 | 128 | 8432.9 | 03 46 40.2628234 | 0.0000006 | +23 29 51.279617 | 0.000022 | 789 | 15 |
| 2456255.71 2012.8967 | 128 | 8432.9 | 03 46 40.2626112 | 0.0000004 | +23 29 51.273089 | 0.000017 | 1184 | 16 |
| 2456480.09 2013.5112 | 128 | 8432.9 | — | — | — | — | — | 14 |
| 2456521.99 2013.6260 | 512 | 8415.9 | 03 46 40.2640676 | 0.0000044 | +23 29 51.239909 | 0.000181 | 48 | 8 |

**Table S4.**

Observations Summary for HII 2147 NW

| Separation from J0347+2339 = 0.288° | | | | $V_{mag}$= 10.9, $B_{mag}$–$V_{mag}$ color = 0.8 (combined light with HII 2147 SE) | | | | |
|---|---|---|---|---|---|---|---|---|
| Observation Date (mid-track) | Band width | Central Frequency | Measured R. A. | R. A. error (one S.D.) | Measured Decl. | Decl. error (one S.D.) | Flux | Map rms noise |
| Julian Date Years | MHz | GHz | h m s | s | ° ' " | " | µJy | µJy |
| 2455926.61 2011.9976 | 128 | 8432.9 | 03 49 06.1250154 | 0.0000039 | +23 46 51.934386 | 0.000270 | 89 | 15 |
| 2455968.50 2012.1120 | 128 | 8432.9 | — | — | — | — | — | 16 |
| 2456010.39 2012.2265 | 128 | 8432.9 | — | — | — | — | — | 15 |
| 2456053.27 2012.3436 | 128 | 8432.9 | 03 49 06.12576225 | 0.0000009 | +23 46 51.9192502 | 0.000045 | 495 | 19 |
| 2456095.16 2012.4581 | 128 | 8432.9 | 03 49 06.1263090 | 0.0000036 | +23 46 51.915510 | 0.000161 | 113 | 16 |
| 2456137.03 2012.5725 | 128 | 8432.9 | — | — | — | — | — | 18 |
| 2456178.94 2012.6870 | 128 | 8432.9 | 03 49 06.1269695 | 0.0000019 | +23 46 51.905948 | 0.000048 | 364 | 12 |
| 2456221.81 2012.8041 | 128 | 8432.9 | 03 49 06.1269222 | 0.0000018 | +23 46 51.900329 | 0.000071 | 148 | 13 |
| 2456304.58 2013.0304 | 128 | 8432.9 | 03 49 06.1266394 | 0.0000041 | +23 46 51.887238 | 0.000157 | 135 | 18 |
| 2456346.47 2013.1451 | 128 | 8432.9 | 03 49 06.1267233 | 0.0000044 | +23 46 51.881440 | 0.000142 | 156 | 20 |
| 2456388.84 2013.2612 | 512 | 8415.9 | 03 49 06.1270808 | 0.0000087 | +23 46 51.875953 | 0.000317 | 43 | 5 |
| 2456472.11 2013.4894 | 512 | 8415.9 | 03 49 06.1281618 | 0.0000026 | +23 46 51.866556 | 0.000085 | 33 | 7 |
| 2456515.00 2013.6069 | 512 | 8415.9 | 03 49 06.1285636 | 0.0000018 | +23 46 51.862185 | 0.000045 | 184 | 9 |
| 2456556.89 2013.7216 | 512 | 8415.9 | — | — | — | — | — | 19 |

**Table S5.**

Observations Summary for HII 2147 SE

| Separation from J0347+2339 = 0.288° | | | | | $V_{mag}$= 10.9, $B_{mag}$–$V_{mag}$ color = 0.8 (combined light with HII 2147 NW) | | | |
|---|---|---|---|---|---|---|---|---|
| Observation Date (mid-track) | Band width | Central Frequency | Measured R. A. | R. A. error (one S.D.) | Measured Decl. | Decl. error (one S.D.) | Flux | Map rms noise |
| Julian Date | MHz | GHz | h m s | s | ° ′ ″ | ″ | µJy | µJy |
| 2455926.61 2011.9976 | 128 | 8432.9 | — | — | — | — | — | 15 |
| 2455968.50 2012.1120 | 128 | 8432.9 | 03 49 06.12751797 | 0.0000025 | +23 46 51.8772440 | 0.000113 | 173 | 16 |
| 2456010.39 2012.2265 | 128 | 8432.9 | 03 49 06.12771442 | 0.0000038 | +23 46 51.8722728 | 0.000163 | 125 | 15 |
| 2456053.27 2012.3436 | 128 | 8432.9 | 03 49 06.12813948 | 0.0000013 | +23 46 51.8680450 | 0.000056 | 276 | 19 |
| 2456095.16 2012.4581 | 128 | 8432.9 | — | — | — | — | — | 16 |
| 2456137.03 2012.5725 | 128 | 8432.9 | 03 49 06.1289993 | 0.0000008 | +23 46 51.860142 | 0.000023 | 791 | 18 |
| 2456178.94 2012.6870 | 128 | 8432.9 | 03 49 06.1291412 | 0.0000007 | +23 46 51.855428 | 0.000026 | 534 | 12 |
| 2456221.81 2012.8041 | 128 | 8432.9 | 03 49 06.1290302 | 0.0000011 | +23 46 51.849802 | 0.000023 | 1157 | 13 |
| 2456304.58 2013.0304 | 128 | 8432.9 | 03 49 06.1285768 | 0.0000037 | +23 46 51.838635 | 0.000237 | 100 | 18 |
| 2456346.47 2013.1451 | 128 | 8432.9 | 03 49 06.1285947 | 0.0000042 | +23 46 51.833105 | 0.000187 | 144 | 19 |
| 2456388.84 2013.2612 | 512 | 8415.9 | 03 49 06.1288710 | 0.0000014 | +23 46 51.828353 | 0.000060 | 90 | 5 |
| 2456472.11 2013.4894 | 512 | 8415.9 | 03 49 06.1297639 | 0.0000007 | +23 46 51.820971 | 0.000032 | 219 | 7 |
| 2456515.00 2013.6069 | 512 | 8415.9 | 03 49 06.1300836 | 0.0000037 | +23 46 51.816703 | 0.000084 | 84 | 9 |
| 2456556.89 2013.7216 | 512 | 8415.9 | 03 49 06.1301307 | 0.0000049 | +23 46 51.812418 | 0.000157 | 119 | 19 |